\documentclass[letterpaper,10pt]{article}
\usepackage[utf8]{inputenc}

\usepackage{color}
\usepackage{xcolor}
\usepackage{graphicx}
\usepackage{tikzscale}
\usepackage{tikz}
\usepackage{pgfplots}
\usetikzlibrary{plotmarks,matrix,chains,scopes,fit,calc,shapes,positioning,decorations,intersections,fit,backgrounds,patterns}
\pgfplotsset{compat=newest} 

\pgfplotsset{plot coordinates/math parser=false}
\pgfplotsset{every axis plot/.append style={solid,line width=1.5pt,mark size=1.5pt,mark options={solid,fill=white}}}
\pgfplotsset{every axis legend/.append style={legend cell align=left,font=\footnotesize}}

\colorlet{42GBd16QAM_color}{blue!80!white}
\colorlet{42GBd64QAM_color}{red!80!black}
\colorlet{42GBd64QAMnoNL_color}{blue!80!white}
\colorlet{64GBd64QAM_color}{gray!50!black}

\pgfplotsset{64GBd64QAM/.style={color=64GBd64QAM_color,solid}}
\pgfplotsset{64GBd16QAM/.style={color=64GBd16QAM_color,dotted}}
\pgfplotsset{42GBd64QAM/.style={color=42GBd64QAM_color,solid}}
\pgfplotsset{42GBd64QAMnoNL/.style={color=42GBd64QAMnoNL_color,dashed}}
\pgfplotsset{42GBd16QAM/.style={color=42GBd16QAM_color,dash dot dot}}

\newlength\FigureWidth
\newlength\FigureHeight
\newlength\FullFigureWidth
\graphicspath{{figures/}}

\pgfplotsset{myLegend/.append style={legend style={font=\footnotesize,at={(0.5,0.98)},anchor=north,align=left,legend columns=3}}}

\newcommand{\AIRn}{\ensuremath{\text{AIR}_n}\xspace}

\usepackage{osameet3} 
\usepackage{amsmath,amssymb}
\usepackage[normalem]{ulem}

\usepackage{xspace}
\usepackage{wrapfig}
\usepackage{caption}

\usepackage[colorlinks=true,bookmarks=false,citecolor=blue,urlcolor=blue]{hyperref} 

\begin{document}

\title{On the Impact of Finite-Length Probabilistic Shaping on Fiber Nonlinear Interference}
\author{Tobias Fehenberger}
\address{ADVA, Fraunhoferstr. 9a, 82152 Martinsried/Munich, Germany\\}
\email{\href{mailto:tfehenberger@adva.com}{tfehenberger@adva.com}}
\begin{abstract} 
The interplay of shaped signaling and fiber nonlinearities is reviewed in the asymptotic and finite-length regime. We present explanations and discuss implications of an optimum shaping length of just a few hundred symbols.
\end{abstract}

\vspace{-4pt}
\section{Introduction}
\vspace{-7pt}

Probabilistic constellation shaping (PCS) is a technique that offers a shaping gain of up to 1.53~dB signal-to-noise ratio (SNR) for the linear additive white Gaussian noise (AWGN) channel \cite{Forney}. In fiber-optic communications, the first use of PCS, to the best of our knowledge, was to reduce the peak-to-average power in OFDM systems via Trellis shaping \cite{HellerbrandTrellisShaping}. The first papers to employ PCS to achieve a shaping gain were published in 2012 \cite{Smith2012JLT_CodedModulation,beygi2012adaptive}. PCS has attracted wide interest since the proposal of probabilistic amplitude shaping (PAS) framework in 2014/15 \cite{GeorgTComm}. The first demonstrations were published the same year \cite{myShapingOFC,ShapingPDP,myShapingPTL}. Since then, countless papers have been written on PCS, and it has been implemented in commercial high-performance digital signal processors (DSPs).

\vspace{-9pt}
\section{Fundamentals of PCS}
\vspace{-7pt}

The success story of PCS has several reasons. The underlying PAS architecture enables a low-complexity integration of PCS into existing coded modulation schemes with off-the-shelf binary forward error correction (FEC). Furthermore, a shaping gain of approx. 1~dB for high-order quadrature amplitude modulation (QAM) formats can be a significant benefit for optimized fiber-optic communication systems. Another important feature of PCS is rate adaptivity, which means that the throughput can be varied by changing the shaping distribution while keeping QAM order and FEC overhead fixed. This allows to dynamically and efficiently utilize the available spectrum.

The processing block that enables PAS is the distribution matcher (DM), which transforms a block of $k$ uniformly distributed input bits into a sequence of $n$ shaped amplitudes of the desired distribution. The initially proposed constant-composition distribution matcher (CCDM) \cite{Schulte2016TransIT_DistributionMatcher} is asymptotically optimal, yet its finite-length rate loss has been shown to be suboptimal \cite{MPDM}, which is why long CCDM blocks would be beneficial. All CCDM algorithms available in the literature are, however, sequential \cite{PASR}. Obviously, the combination of long blocks and sequential algorithms is challenging for real-time implementation. Research effort was put into finding algorithms that had the lowest rate loss at a fixed block length $n$ or offered some implementation benefit.

\vspace{-9pt}
\section{Nonlinear Interference for PCS}
\vspace{-7pt}
For the nonlinear fiber channel, PCS has been examined in theory, simulations, and experiments \cite{myJLT,JulianJLT,ESS_Karim,ESS_Sebastiaan}. A decrease in effective SNR after DSP compared to uniform signaling is observed due to the increased kurtosis of the shaped constellation, yet this only marginally reduces the shaping gain that could be achieved over a linear channel \cite{myJLT}. Most studies of the impact of PCS on NLI do not focus on the above mentioned finite-length aspects and thus use a simplified PAS setup. A common PAS emulation technique is to simply draw the QAM symbols according to the desired distribution without considering DM implementation aspects at all. A slightly more realistic approach is to consider very long CCDM sequences of several thousand or tens of thousands amplitudes, neglecting whether such long blocks are implementable in practice.

\begin{wrapfigure}{R}{0.7\textwidth}
  \vspace*{-\baselineskip}
  \centering
  \definecolor{mycolor1}{rgb}{0.00000,0.44700,0.74100}%
\definecolor{mycolor2}{rgb}{0.85000,0.32500,0.09800}%
\definecolor{mycolor3}{rgb}{0.92900,0.69400,0.12500}%
\definecolor{mycolor4}{rgb}{0.49400,0.18400,0.55600}%
\definecolor{mycolor5}{rgb}{0.46600,0.67400,0.18800}%
\begin{tikzpicture}[font=\small]


\begin{axis}[%
width=0.35\textwidth,
height=\FigureHeight,
xmin=10,
xmax=5000,
xlabel={CCDM block length $n$},
ymin=18.5,
ymax=22,
ylabel={Effective SNR [dB]},
xmode=log,
xmajorgrids,
ymajorgrids,
extra x ticks={5000},
extra x tick style={xticklabel=\hspace{-1em}$5\!\cdot\!10^3$},
legend style={at={(0.99,0.99)}, anchor=north east},
clip mode=individual,
name=plot1,
inner sep=2pt,
]


\addplot[42GBd64QAM]
  table[row sep=crcr]{%
10  19.60892029  \\
50  19.48604469  \\
100  19.40795393  \\
200  19.32237693  \\
500  19.16954758  \\
1000  19.02666493  \\
2000  18.91938521  \\
3000  18.86808549  \\
4000  18.8440601  \\
5000  18.83558265  \\
};

\addplot[42GBd64QAMnoNL]
  table[row sep=crcr]{%
10  21.74  \\
50  21.74  \\
100  21.74  \\
200  21.73  \\
500  21.75  \\
1000  21.74  \\
2000  21.75  \\
3000  21.75  \\
4000  21.74  \\
5000  21.75  \\
};

\addplot[gray,dotted]
  table[row sep=crcr]{%
10  19.12  \\
50  19.12  \\
100  19.12  \\
200  19.12  \\
500  19.12  \\
1000  19.12  \\
2000  19.12  \\
3000  19.12  \\
4000  19.12  \\
5000  19.12  \\
};




\node[anchor=west] at (30,21.5) {\textcolor{42GBd64QAMnoNL_color}{without NLI}};
\node[anchor=west,rotate=-9] at (30,19.9) {\textcolor{42GBd64QAM_color}{with NLI}};
\node[anchor=north west] at (10,19.12) {\textcolor{gray}{uniform}};






\coordinate (insetn20) at (30,19.25);
\coordinate (insetn5000) at (1700,19.3);


\end{axis}%

\end{tikzpicture}%

  \definecolor{mycolor1}{rgb}{0.00000,0.44700,0.74100}%
\definecolor{mycolor2}{rgb}{0.85000,0.32500,0.09800}%
\definecolor{mycolor3}{rgb}{0.92900,0.69400,0.12500}%
\definecolor{mycolor4}{rgb}{0.49400,0.18400,0.55600}%
\definecolor{mycolor5}{rgb}{0.46600,0.67400,0.18800}%
\begin{tikzpicture}[font=\small]

\begin{axis}[%
width=0.35\textwidth,
height=\FigureHeight,
xmin=10,
xmax=5000,
xlabel={CCDM block length $n$},
ymin=10.8,
ymax=11.4,
ylabel={\AIRn [bit/4D-symbol]},
xmode=log,
xmajorgrids,
ymajorgrids,
extra x ticks={5000},
extra x tick style={xticklabel=\hspace{-1em}$5\!\cdot\!10^3$},
legend style={at={(0.99,0.99)}, anchor=north east},
clip mode=individual,
name=plot2,
at=(plot1.south east), anchor=left of south west,
inner sep=2pt,
]

\addplot[42GBd64QAM]
  table[row sep=crcr]{%
10 10.63955169 \\
50 11.07167111 \\
100 11.13069424 \\
200 11.15697087 \\
500 11.16080754 \\
1000 11.14437322 \\
2000 11.1291782 \\
3000 11.12116516 \\
4000 11.11637905 \\
5000 11.11553206 \\
};

\addplot[42GBd64QAMnoNL]
  table[row sep=crcr]{%
10 10.8236 \\
50 11.2036 \\
100 11.2736 \\
200 11.3136 \\
500 11.3436 \\
1000 11.3546 \\
2000 11.3616 \\
3000 11.3642 \\
4000 11.3656 \\
5000 11.3664 \\
};

\node[anchor=west] at (200,11.3) {\textcolor{42GBd64QAMnoNL_color}{without NLI}};
\node[anchor=west] at (200,11.1) {\textcolor{42GBd64QAM_color}{with NLI}};

\end{axis}%

\end{tikzpicture}%

  \vspace*{-\baselineskip}
  \caption{\label{fig:SNR_GMI_n} Effective SNR after DSP (left) and \AIRn{} (right) vs. $n$ for shaped 64QAM. A regular nonlinear fiber setup (solid red) and a linearized fiber without NLI (dashed blue) are considered. The SNR for uniform 64QAM is shown as reference.}
  \vspace*{-\baselineskip}
\end{wrapfigure}
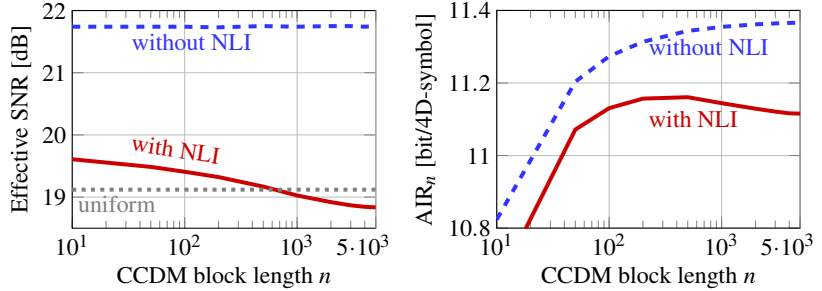

In the first paper to report on the finite-length behavior of PCS with a CCDM \cite{ESS_Karim}, a block-length dependence of SNR was found in simulations, which was later confirmed in experiments \cite{ESS_Sebastiaan}. The SNR is inversely proportional to $n$, i.e., short blocks mitigate NLI and thus give higher SNR than long blocks. Taking into account the finite-length rate loss, the achievable information rate (AIR) \AIRn for length-$n$ DMs \cite{MPDM} is maximized at a finite $n$. This is in stark contrast to a linear channel where SNR is independent of $n$ and AIR increases with $n$ because longer blocks generally lead to lower rate loss. For multi-span WDM fiber simulations \cite{myOFC2020}, Fig.~\ref{fig:SNR_GMI_n} shows an SNR decrease by 0.8~dB between $n=10$ and $n=5000$ as well as an AIR optimum at approximately 500~symbols. When NLI is turned off in the simulations, the SNR becomes independent of $n$ and \AIRn{} is increasing with $n$. 


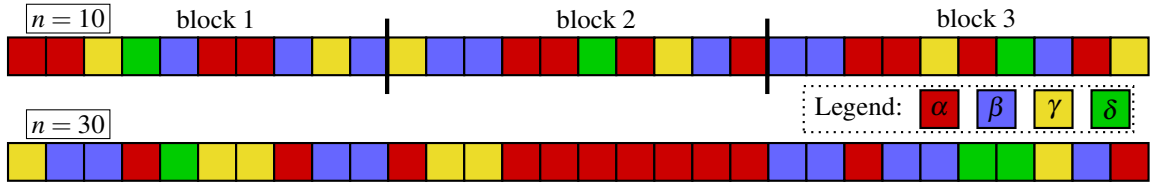
\begin{figure}[!h]
  \vspace*{-0.9\baselineskip}
  \centering
  \begin{tikzpicture}[%
block/.style={rectangle,thick,draw,inner sep=2pt, minimum width=5mm, minimum height=5mm,black, anchor=west,align=center},
]
\colorlet{color0}{red!80!black}
\colorlet{color1}{blue!60!white}
\colorlet{color2}{yellow!90!black}
\colorlet{color3}{green!80!black}

\node[draw,black,anchor=west,inner sep=2pt] at (0.25,0.5) {$n=10$};
\node[block,fill=color0] at (0. ,0) {};
\node[block,fill=color0] at (0.5,0) {};
\node[block,fill=color2] at (1. ,0) {};
\node[block,fill=color3] at (1.5,0) {};
\node[block,fill=color1] at (2. ,0) {};
\node[block,fill=color0] at (2.5,0) {};
\node[block,fill=color0] at (3. ,0) {};
\node[block,fill=color1] at (3.5,0) {};
\node[block,fill=color2] at (4. ,0) {};
\node[block,fill=color1] at (4.5,0) {};
\node[block,fill=color2] at (5. ,0) {};
\node[block,fill=color1] at (5.5,0) {};
\node[block,fill=color1] at (6. ,0) {};
\node[block,fill=color0] at (6.5,0) {};
\node[block,fill=color0] at (7. ,0) {};
\node[block,fill=color3] at (7.5,0) {};
\node[block,fill=color0] at (8. ,0) {};
\node[block,fill=color2] at (8.5,0) {};
\node[block,fill=color1] at (9. ,0) {};
\node[block,fill=color0] at (9.5,0) {};
\node[block,fill=color1] at (10. ,0) {};
\node[block,fill=color1] at (10.5,0) {};
\node[block,fill=color0] at (11. ,0) {};
\node[block,fill=color0] at (11.5,0) {};
\node[block,fill=color2] at (12. ,0) {};
\node[block,fill=color0] at (12.5,0) {};
\node[block,fill=color3] at (13. ,0) {};
\node[block,fill=color1] at (13.5,0) {};
\node[block,fill=color0] at (14. ,0) {};
\node[block,fill=color2] at (14.5,0) {};

\draw[ultra thick] (5,-0.5) -- (5,0.5);

\draw[ultra thick] (10,-0.5) -- (10,0.5);

\node[anchor=center] at (2.75,0.5) {block 1};
\node[anchor=center] at (7.75,0.5) {block 2};
\node[anchor=center] at (12.75,0.5) {block 3};


\node[draw,black,anchor=west,inner sep=2pt] at (0.25,-0.9) {$n=30$};
\node[block,fill=color2] at (0. ,-1.4) {};
\node[block,fill=color1] at (0.5,-1.4) {};
\node[block,fill=color1] at (1. ,-1.4) {};
\node[block,fill=color0] at (1.5,-1.4) {};
\node[block,fill=color3] at (2. ,-1.4) {};
\node[block,fill=color2] at (2.5,-1.4) {};
\node[block,fill=color2] at (3. ,-1.4) {};
\node[block,fill=color0] at (3.5,-1.4) {};
\node[block,fill=color1] at (4. ,-1.4) {};
\node[block,fill=color1] at (4.5,-1.4) {};
\node[block,fill=color0] at (5. ,-1.4) {};
\node[block,fill=color2] at (5.5,-1.4) {};
\node[block,fill=color2] at (6. ,-1.4) {};
\node[block,fill=color0] at (6.5,-1.4) {};
\node[block,fill=color0] at (7. ,-1.4) {};
\node[block,fill=color0] at (7.5,-1.4) {};
\node[block,fill=color0] at (8. ,-1.4) {};
\node[block,fill=color0] at (8.5,-1.4) {};
\node[block,fill=color0] at (9. ,-1.4) {};
\node[block,fill=color0] at (9.5,-1.4) {};
\node[block,fill=color1] at (10. ,-1.4) {};
\node[block,fill=color1] at (10.5,-1.4) {};
\node[block,fill=color0] at (11. ,-1.4) {};
\node[block,fill=color1] at (11.5,-1.4) {};
\node[block,fill=color1] at (12. ,-1.4) {};
\node[block,fill=color3] at (12.5,-1.4) {};
\node[block,fill=color3] at (13. ,-1.4) {};
\node[block,fill=color2] at (13.5,-1.4) {};
\node[block,fill=color1] at (14. ,-1.4) {};
\node[block,fill=color0] at (14.5,-1.4) {};

\node[inner sep=3pt] (legend00) at (11.2,-0.7) {Legend:};
\node[block,fill=color0] (legend0) at (12,-0.7) {$\alpha$};
\node[block,fill=color1] (legend1) at (12.75,-0.7) {$\beta$};
\node[block,fill=color2] (legend2) at (13.5,-0.7) {$\gamma$};
\node[block,fill=color3] (legend3) at (14.25,-0.7) {$\delta$};

\node[fit=(legend0)(legend00)(legend1)(legend2)(legend3), thick, draw, dotted,inner sep=1pt] (box) {};

\end{tikzpicture}

  \vspace*{-0.6\baselineskip}
  \caption{\label{fig:CCDM_block_example} Concatenation of three CCDM blocks of length $n=10$ each (top) in comparison to a single CCDM block of length $n=30$ (bottom). Both cases give the same average distribution, but their temporal properties differ.}
  \vspace*{-\baselineskip}
\end{figure}

The NLI mitigation at short block lengths can be explained as follows for CCDM sequences \cite{myCCJLT2019}. We keep the amplitude distribution  fixed such that a concatenation of CCDM blocks of varying $n$ always has the same average distribution. The only difference is in how many sub-blocks (having constant composition) the compound sequence consists of. The reason for the NLI mitigation must thus lie in temporal properties that are introduced by short-length PCS but are not necessarily present for long blocks. An illustrative explanation is presented in Fig.~\ref{fig:CCDM_block_example} for the distribution $[0.4, 0.3, 0.2, 0.1]$ of four shaped amplitudes $[\alpha,\beta,\gamma,\delta]$. The compound sequence comprising 30 amplitudes is either generated by concatenating three blocks pf length $n=10$ each or from a single CCDM with $n=30$. For $n=10$, each of the three blocks must, for example, contain the symbol $\delta$ once, which imposes a certain temporal structure in the overall compound sequence that is not present for $n=30$. As shown in Fig.~\ref{fig:CCDM_block_example} for $n=30$, the second and third $\delta$ are intra-block neighbors, which is not possible for $n=10$. We conclude that short-length CCDM introduces a temporal structure that limits the clustering of identical symbols, which in turn leads to NLI mitigation. 

\vspace{-9pt}
\section{Implications and Open Questions}
\vspace{-7pt}
With such a NLI mitigation, the ``longer is better'' paradigm of PCS is apparently not true for the nonlinear fiber channel. This is particularly beneficial for hardware implementation as power consumption and latency requirements can become less stringent. The AIR, however, is relatively insensitive to using too long blocks (see Fig.~\ref{fig:SNR_GMI_n}), so the performance loss compared to the optimum is small when using too large block lengths.

As it is the temporal structure that leads to NLI mitigation, any DSP block that modifies the transmit sequence can significantly reduce this benefit, and interleaving is such an operation. In a modern DSP architecture such as in 400ZR \cite{400ZR}, two interleaving stages exist. A bit interleaver is placed between the inner and outer FEC to break up any error bursts from unsuccessfully decoding an inner FEC codeword. Additionally, a symbol interleaver is placed after QAM mapping to achieve polarization and phase diversity, which means that a FEC word is spread over both polarizations and also over time such that it experiences different amounts of phase noise. The shuffling of the bit interleaver can be undone by de-interleaving after the inner FEC encoding and re-applying the interleaving before the inner FEC decoding \cite{myOFC2020}. This ensures that burst errors of an inner codeword are still spread over several outer FEC blocks, yet the temporal structure required for NLI mitigation remains intact. While this is in principle feasible, it remains unclear how practical such an approach is and whether it has any other drawback. Regarding symbol interleaving, there is no straightforward way of keeping the diversity due to scrambling while achieving the NLI mitigation by short PCS. Thus, it remains an open question whether NLI mitigation or polarization and phase diversity have a higher overall impact on the system performance. 


\vspace{-19pt}
\begin{thebibliography}{99}
\vspace{-4pt}
\footnotesize
\setlength{\itemindent}{-12pt}%

\bibitem{Forney} G.~D.~Forney \textit{et al.}, “Efficient modulation for band-limited channels,” IEEE JSAC, 1984.

\bibitem{HellerbrandTrellisShaping}
S.~Hellerbrand \textit{et al.}, “Trellis shaping for reduction of the peak-to-average power ratio in coherent optical OFDM systems,” OFC, 2009. 

\bibitem{Smith2012JLT_CodedModulation} B.~P. Smith and F.~R. Kschischang, ``{A pragmatic coded modulation scheme for high-spectral-efficiency fiber-optic \dots},'' JLT, 2012.

\bibitem{beygi2012adaptive}
L.~Beygi \textit{et al.}, ``{Adaptive coded modulation for
  nonlinear fiber-optical channels},'' IEEE GLOBECOM, 2012.

\bibitem{GeorgTComm} G. Böcherer \textit{et al.}, ``Bandwidth efficient and rate-matched low-density parity-check coded modulation,'' IEEE Trans. Comm., 2015.

\bibitem{myShapingOFC} T.~Fehenberger \textit{et al.}, “LDPC coded modulation with probabilistic shaping for optical fiber systems,” OFC, 2015.

\bibitem{ShapingPDP} F.~Buchali \textit{et al.}, “Experimental demonstration of capacity increase and rate-adaptation by probabilistically shaped \dots,” ECOC, 2015.

\bibitem{myShapingPTL} T.~Fehenberger \textit{et al.}, “Sensitivity gains by mismatched probabilistic shaping for optical communication
systems,” PTL, 2016.

\bibitem{Schulte2016TransIT_DistributionMatcher} P.~Schulte and G.~Böcherer, ``Constant composition distribution matching,'' IEEE Trans. IT, 2016.


\bibitem{MPDM} T.~Fehenberger \textit{et al.},  ``{Multiset-partition distribution matching},'' IEEE Trans. Comm., 2019.

\bibitem{PASR} T.~Fehenberger \textit{et al.},  ``{Parallel-amplitude architecture and subset ranking for fast distribution matching},'' IEEE Trans. Comm., 2020.

\bibitem{myJLT} T.~Fehenberger \textit{et al.}, ``On probabilistic shaping of quadrature amplitude modulation for the nonlinear fiber channel,'' JLT, 2016.

\bibitem{JulianJLT} J.~Renner \textit{et al.}, “Experimental comparison of probabilistic shaping methods for unrepeated fiber transmission,” JLT, 2017.


\bibitem{ESS_Karim}
A.~Amari \textit{et al.},  ``Introducing enumerative sphere shaping for optical communication systems with short ...,'' JLT, 2019.

\bibitem{ESS_Sebastiaan} S.~Goossens \textit{et al.},  ``First experimental demonstration of probabilistic enumerative sphere shaping in ...,'' arXiv:1908.00453v2, 2019.

\bibitem{myOFC2020} T.~Fehenberger \textit{et al.}, ``{Mitigating fiber nonlinearities by short-length probabilistic shaping},'' OFC, 2020.

\bibitem{myCCJLT2019} T.~Fehenberger \textit{et al.}, ``{Analysis of nonlinear fiber interactions for finite-length constant-composition sequences},'' JLT, 2019.





\bibitem{400ZR} I.~Lyubomirsky \textit{et al.}, ``400GBASE-ZR PCS/PMA baseline proposal,'' IEEE P802.3cn Task Force Meeting, 2019. 




\end{thebibliography}
\end{document}